\documentclass[prl,english,superscriptaddress,twocolumn]{revtex4}
\usepackage{amsmath,amssymb,amsfonts,mathrsfs}
\usepackage{amsthm}
\usepackage{CJK}
\usepackage{bm}
\usepackage{babel}
\usepackage{graphicx}

\begin{document}

\title{Realizing and manipulating space-time inversion symmetric topological
semimetal bands with superconducting quantum circuits}
\author{Xinsheng Tan}
\affiliation{National Laboratory of Solid State Microstructures, School of Physics,
Nanjing University, Nanjing 210093, China}
\author{Y. X. Zhao}
\affiliation{Department of Physics and Center of Theoretical and Computational Physics,
The University of Hong Kong, Pokfulam Road, Hong Kong, China}
\affiliation{Max-Planck-Institute for Solid State Research, D-70569 Stuttgart, Germany}
\author{Qiang Liu}
\affiliation{National Laboratory of Solid State Microstructures, School of Physics,
Nanjing University, Nanjing 210093, China}
\author{Guangming Xue}
\affiliation{National Laboratory of Solid State Microstructures, School of Physics,
Nanjing University, Nanjing 210093, China}
\author{Haifeng Yu}
\email[]{hfyu@nju.edu.cn}
\affiliation{National Laboratory of Solid State Microstructures, School of Physics,
Nanjing University, Nanjing 210093, China}
\affiliation{Synergetic Innovation Center of Quantum Information and Quantum Physics,
University of Science and Technology of China, Hefei, Anhui 230026, China,}
\author{Z. D. Wang}
\email[]{zwang@hku.hk}
\affiliation{Department of Physics and Center of Theoretical and Computational Physics,
The University of Hong Kong, Pokfulam Road, Hong Kong, China}
\author{Yang Yu}
\email[]{yuyang@nju.edu.cn}
\affiliation{National Laboratory of Solid State Microstructures, School of Physics,
Nanjing University, Nanjing 210093, China}
\affiliation{Synergetic Innovation Center of Quantum Information and Quantum Physics,
University of Science and Technology of China, Hefei, Anhui 230026, China,}

\begin{abstract}
We have experimentally realized novel space-time
inversion (P-T) invariant  $\mathbb{Z}_2$-type topological semimetal-bands,  
via an 
analogy between the momentum space and a controllable parameter space
in superconducting quantum circuits. By measuring the whole energy spectrum
of system, we imaged clearly an exotic tunable gapless band structure of
topological semimetals.  
Two topological quantum phase transitions from a topological semimetal to
two kinds of insulators can be manipulated by continuously tuning the
different parameters in the experimental setup, one of which captures the $%
\mathbb{Z}_2$ topology of the $PT$ semimetal via merging a pair of nontrivial $\mathbb{Z%
}_2$ Dirac points.  Remarkably, the topological robustness was
demonstrated unambiguously, by adding a perturbation that breaks only the
individual $T$ and $P$ symmetries but keeps the joint $PT$ symmetry.  
In contrast, when another kind of $PT$-violated perturbation is introduced,
a topologically trivial insulator gap is fully opened.
\end{abstract}

\maketitle

Symmetry and topology, as the two fundamentally important concepts in
physics and mathematics, have not only manifested themselves in science, but
also provided us profound understanding of arresting natural phenomena.
Recently, topological gapless systems, such as 
Weyl semimetals \cite%
{Xu-Science15,LvWeyl,Lu-Science15,XGWan-PRB11,Zhao-Wang-PRL15} and a variety
of Dirac semimetals \cite%
{LiuDirac,Neupane-NC15,Nagaosa-nc14,KimDiracline,YuDiracline} as well as $%
\mathbb{Z}_{2}$ topological metals/semimetals~\cite%
{ZhaoWangPRL13,Furusaki-Z2-PRB14,zhao_wang_PRL_16}, have 
significantly stimulated research interest. Analogous to that in gapped topological
systems such as topological insulators and superconductors, the discrete
symmetry that is rather robust against symmetry-preserved perturbations can
enrich the topological physics of gapless systems as well. As is known, the
discrete time-reversal ($T$), space-inversion ($P$), and charge-conjugate ($C
$) symmetries are fundamental and intriguing in nature. For examples, in
high energy physics, any local quantum field theory must preserve the joint $%
CPT$ symmetry, which is required by the unitarity and Lorentz invariance of
the theory, and the source of $CP$ violation still remains as one of seminal
mysteries in the Standard Model. While in condensed matter systems, it is
ubiquitous that $P$, $T$ and $C$ put the constraints on band structures and
lead to new topological classifications of band theories \cite%
{Volovik:book,Hasan:2010kx,Qi:2011vn}. Among various combinations of $P$, $T$
and $C$, the joint $PT$ symmetry actually inverses the space-time
coordinates $x^{\mu }\rightarrow -x^{\mu }$ with $\mu =0,1,2,3$ and $x^{0}=t$%
, and therefore evidences itself to be fundamental and significant in
physics. Very recently, a theory of $PT$-invariant topological gapless bands
has rigorously been established~\cite{Zhao-Schnyder-Wang-PRL}, through
revealing a profound connection between the $PT$ symmetry 
and an elegant $KO$ theory of algebraic topology~\cite{Atiyah-KR}. 
On the other hand, it is noted that artificial superconducting quantum
circuits possess high controllability~\cite%
{makhlin,vion,nakamura,Gritsev,Yu-Science02}, providing a powerful and ideal
tool to quantum-simulate and explore novel quantum systems~\cite%
{nist,roushan,Lupascu-PRL2015,you2011atomic,Tan-PRL14}, including topological ones.

In this Letter, we have 
realized experimentally the novel $PT$ symmetry-protected topological
semimetal-bands that represent a gapless spectrum on a square-lattice, via
an analogy between the momentum space and a controllable parameter space in
superconducting quantum circuits. By measuring the whole energy spectrum of
our system, we have imaged clearly an exotic tunable gapless band structure
of topological semimetals, shown as nontrivial $\mathbb{Z}_{2}$-type Dirac
points in momentum space. The two new distinct quantum phase transitions
from a topological semimetal to two different insulators can be manipulated
by continuously tuning the different parameters in the simulated effective
Hamiltonian, particularly one of which exhibits the $\mathbb{Z}_{2}$ topology in the $PT$
semimetal via merging a pair of nontrivial $\mathbb{Z}_{2}$ Dirac points.
Furthermore, to demonstrate unambiguously the topological robustness of $PT$
symmetry, a perturbation that breaks only the individual $T$ and $P$
symmetries is intentionally added, with the joint $PT$ symmetry being still
preserved. It is verified by experimental date that the Dirac points of the
topological semimetal-bands still present under such perturbations, though the point positions and
the band pattern are changed drastically. In a sharp contrast, when another
kind of perturbation is added to break the $PT$ symmetry in our experiment,
the energy gap is fully opened and the Dirac points disappear completely, showing the essential role of $PT$ symmetry underlying the topological robustness.
All of these illustrate convincingly the topological protection of $PT$
semimetals. Notably, the present work is the first experimental realization
and manipulation of fundamental space-time inversion symmetric topological
semimetal-bands (without individual $T$ and $P$ symmetries) in nature, which
opens a window for simulating and manipulating topological quantum
matter.

The physical manifestation of $PT$ symmetry in band
theories can simply be seen from the commutation relation as $[\hat{A},H]=0$%
, where $H$ is the system Hamiltonian, and the joint $PT$ symmetry is
represented by an anti-unitary operator $\hat{A}$~\cite{Notes-PT-locality}. When $\hat{A}^2=1$, the
topological classification of band-crossing points in two-dimensional band
structures corresponds to the reduced $KO$ group, $\widetilde{KO}%
(S^{1})\cong \mathbb{Z}_2$, which implies that there exist band-crossing
points having nontrivial $\mathbb{Z}_2$ topological charges in two
dimensions~\cite{Zhao-Schnyder-Wang-PRL}. Although the $KO$ theory of algebraic topology seems to be rather abstruse
for most physicists, the predicted topological band crossing points can be realized in a simple but representative
dimensionless Hamiltonian, which is explicitly given by~\cite%
{Zhao-Schnyder-Wang-PRL} 
\begin{equation}
H(k)=\sin k_{x}\sigma _{2}+(\lambda \pm \cos k_{y})\sigma _{3}
\label{Lattice-model}
\end{equation}
with the $PT$ being denoted by $\hat{A}=\sigma_3\hat{\mathcal{K}}$, where $%
\sigma_j$ is the $j$th Pauli matrix, and $\hat{\mathcal{K}}$ denotes the
complex conjugate operation. When $-1<\lambda<1$, the model %
\eqref{Lattice-model}, which describes actually a topologically nontrivial
spin(1/2)-orbital quantum system in two dimension, has four band-crossing
points possessing the $PT$-protected $\mathbb{Z}_2$ ($\nu_{\mathbb{Z}_2}=1$) topological charges.
It is noted that although the model \eqref{Lattice-model} has both $\hat{P}%
=\sigma_3\hat{i}$ and $\hat{T}=\hat{\mathcal{K}}\hat{i}$ symmetries with $%
\hat{i}$ being the inversion of the wave vector $k$, the topological
stability of these band-crossing points merely requires the joint $PT$
symmetry according to the $PT$ invariant topological band theory, namely,
the $T/P$-symmetry is allowed to be broken individually while the $PT$
topological protection still remains.

\begin{figure}[tbp]
\includegraphics[scale=0.5]{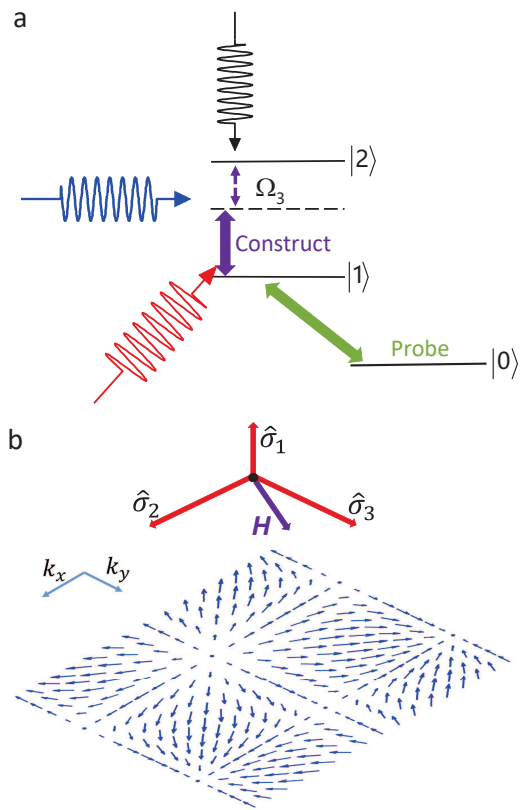} 
\caption{Experimental scheme for the realization of the lattice Hamiltonian.
a, States $|2\rangle $ and $|1\rangle $ of a transmon are used as the energy
levels of an artificial spin-1/2 particle, whose three components may be denoted by the three Pauli matrices $\hat{\sigma}_{1,2,3}$. $|0\rangle $ is chosen as an
ancillary level to probe the eigenvalues of a Hamiltonian. Microwaves with various frequencies, phases, and amplitude are
applied for the construction of
a semimetal Hamiltonian and circuit QED readout, respectively \cite{Notes-mapping}. b, The constructed Hamiltonian is implemented with modulation of microwave amplitude,
frequency, and phase, mapping to the momentum space of a square lattice. }
\label{Scheme}
\end{figure}

Experimental demonstration of this new kind of symmetry protected
topological gapless band will significantly deepen our understanding of
topological quantum matter. However, there are several big challenges that
hinder the realization and investigation of the topological properties of
this kind of Hamiltonians in real condensed matter systems. The first is how
to synthesize the materials with a designated Hamiltonian. Secondly, even if
one is fortune enough to have such kind of real materials, it seems
extremely hard to tune the parameters continuously for studying fruitful
topological properties including various topological quantum phase
transitions. Moreover, it is quite difficult in experiments to directly
image the whole momentum-dependent electronic energy spectrum of a bulk
condensed matter system, noting that only a part of electronic spectra (or information of Fermi surfaces/points) may be inferred from
the angle-resolved photoemission spectroscopy data (or quantum oscillation measurements). 
Therefore, it is imperative and
important as well as significant to use artificial quantum systems like
superconducting quantum circuits to simulate $H(k)$ faithfully and to
explore topological properties of the system. Below we will realize the
Hamiltonian of Eq.\eqref{Lattice-model} in the parameter (analogous
to the momentum) space via implementing a kind of fully controllable quantum
superconducting circuits, 
such that the band structure can directly be measured over the whole first
Brillouin zone (BZ) of square lattices, enabling us to demonstrate the
unique topological nature of the corresponding semimetal-bands and to
clearly visualize some crucial properties. 

The superconducting quantum circuits used in our experiment consist of a
superconducting transmon qubit embedded in a three dimensional aluminium
cavity~\cite{wallraff,koch,Blais,devoret_3d,shankar_3d,wang_3d}. The transmon qubit, which is  composed of
a single Josephson junction and two pads (250 $\mu $m $\times $ 500 $\mu $%
m), is patterned using standard e-beam lithography, followed by double-angle
evaporation of aluminium on a 500 $\mu $m thick silicon substrate. The
thicknesses of the Al film are 30nm and 80nm, respectively. The chip is
diced into 3 mm $\times $ 6.8 mm size to fit into the 3D rectangular
aliminium cavity with the resonance frequency of TE101 mode 9.053 GHz. The
whole sample package is cooled in a dilution refrigerator to a base
temperature 30 mK. The dynamics of the system is identical to an artificial
atom located in a cavity which has been extensively discussed as a circuit
QED\cite{You,wallraff,Blais}. We designed the energy level of the transmon qubit to let the system
work in the dispersive region. The quantum states of the transmon qubit can
be controlled by microwaves. Inphase quadrature (IQ) mixers combined with 1 GHz arbitrary wave
generator (AWG) are used to adjust the amplitude, frequency, and phase of
microwave pulses. To read out qubit states, we use ordinary microwave
heterodyne setup. The output microwave is pre-amplified by HEMT at 4 K stage
in the dilution refrigerator and further amplified by two low noise
amplifiers in room temperature. The microwave is then heterodyned into 50
MHz and collected by ADCs. The readout is performed with so called
\textquotedblleft high power readout" scheme \cite{Reed_readout}. By sending in a strong
microwave on-resonance with the cavity, the transmitted amplitude of the
microwave reflects the state of the transmon due to the non-linearity of the
cavity QED system\cite{Notes-setup}.

According to the circuit QED theory, the coupled transmon qubit and cavity
exhibit anharmonic multiple enengy levels. In our experiments, we use the
lowest three energy levels, as shown in Fig.\ref{Scheme}a, namely, $%
|0\rangle $, $|1\rangle ,$ and $|2\rangle $. The two states $|2\rangle $ and 
$|1\rangle $ behave as an artificial spin-1/2 particle, whose three components may be denoted by the three Pauli matrices $\sigma_{1,2,3}$ which can couple
with the microwave fields. $|0\rangle $ is chosen as an ancillary level to probe
the energy spectrum of the simulated system. The transition frequencies
between different energy levels are $\omega _{10}/2\pi =$ 7.17155 GHz, $%
\omega _{21}/2\pi =$ 6.8310 GHz, respectively, which are independently
determined by saturation spectroscopies. The energy relaxation time of the qubit is $T_1\sim$ 15$\mu s$, the dephasing time is $T^*_2\sim$ 4.3$\mu s$. When we apply microwave drive along 
$x$, $y$, and $z$ directions, the effective Hamiltonian of the qubit in the
rotating frame (Fig.\ref{Scheme}b) may be written as ($\hbar =1$ for brevity)

\begin{equation}
\hat{H}=\sum_{i=1}^{3}\Omega _{i}\sigma _{i}/2,
\end{equation}%
where $\Omega _{1}$ $(\Omega _{2})$ corresponds to the frequency of Rabi
oscillations along X (Y) axis on the Bloch sphere, which is continuously
adjustable by changing the amplitude and phase of microwave applied to the
system. $\Omega _{3}=\omega _{21}-\omega ,$ is determined by the detuning
between the system energy level spacing $\omega _{21}$ and microwave
frequency $\omega $. By carefully designing the waveform of AWG, we can
control the frequency, amplitude, and phase of microwave. In our
experiment, we first calibrated the parameters $\Omega _{1},$ $\Omega _{2},$
and $\Omega _{3}$ using Rabi oscillations and Ramsey fringes, and then
designed the microwave amplitude, frequency and phase to let $\Omega _{1}=0,$
$\Omega _{2}(k_{x})=\Omega \sin k_{x},$ $\Omega _{3}(k_{y})=\lambda \Omega
+\Omega \cos k_{y}$, with $\Omega =$ 10 MHz being chosen as the energy unit.

\begin{figure}
	\includegraphics[scale=0.28]{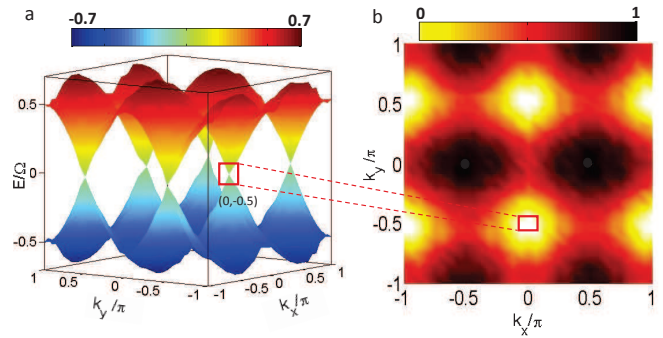} 
	\caption{ Measured energy
spectrum of a typical space-time inversion invariant topological semimetal.
a, Three-dimensional plot of the band structure of spectroscopy measurement.
By tuning the driving amplitude, frequency, and phase gradually, we image
the band structure of the system in the momentum space point by point. b,
Magnitude of energy gap obtained from direct measurements of the energy
spectrum of the system as function of $k_{x}$ and $k_{y}$ in the first BZ.
Four nontrivial $\mathbb{Z}_{2}$-type Dirac points located inside the bright
regions can be observed at $(0,\pm \protect\pi /2)$, $(\protect\pi ,\pm 
\protect\pi /2)$, in a full agreement with the theoretical prediction. \label{Spectrum} }
\end{figure}

Exploiting the analogy between the above parameter space of our system and the k-space of a lattice Hamiltonian system, 
 we now have Eq.\eqref{Lattice-model} exactly. It is
worth to mention that $\lambda $ plays a crucial role in the realization of the $PT$ invariant topological phase transition. To examine the band
structure, we first set $\lambda =0$ and measured the entire energy spectrum
of the system over the first BZ, as shown in Fig. \ref{Spectrum}. In our
experiment, for a given ($k_{x},$ $k_{y})\in [-\pi ,\pi )\times [-\pi,\pi)$, 
we actually measured the resonant peak of microwave absorption, and
determined the frequency of the resonant peak as a function of $k_{x}$ and $%
k_{y}$ \cite{Notes-mapping}. 
A key feature of the $PT$ invariant topological semimetal, which is the
existence of nontrivial $\mathbb{Z}_2$-type Dirac points yielded by crossing
bands, is clearly seen in Fig.\ref{Spectrum}b. These are the directly imaged
Dirac cones in the experiments, indicating that we have successfully
realized the topological semimetal that preserves the $PT$ symmetry. In
addition, the positions of the Dirac points (Fig.\ref{Spectrum}a) locate at (%
$\pi$, $\pm\pi/2$) and ($0$, $\pm\pi /2$), agreeing well with the
theoretical calculation of Eq.\eqref{Lattice-model} with $\lambda=0$.

\begin{figure}[tbp]
\includegraphics[scale=0.28]{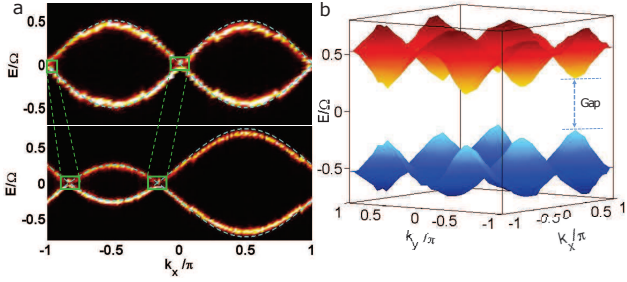} 
\caption{Symmetry-related topological features of the Dirac points for two
different but representative kinds of perturbations. a, When $H_{1}^{\prime
}=\protect\eta \protect\sigma _{2}$ is added with $\protect\eta =0.5$ in
unit of $\Omega $, which breaks both $T$ and $P$ but preserves the $PT$
symmetry, Dirac-like points still exist, though the gapless point positions
are shifted (marked by the green square) and the band pattern is distorted
drastically, showing the robust of the topological nature protected by the $PT$ symmetry. Top and bottom panels correspond respectively to the cases of $\eta =0$ and $\eta =0.5$ on the plane of $k_y=\pi/2$. The bright yellow and dashed green
lines denote the experimental data and theoretical calculations from Eq.(\ref{Lattice-model}) with $H'_1$ being added,
respectively. b, Whenever the $PT$ symmetry is broken by adding the term $H_{2}^{\prime }=\protect\varepsilon \protect\sigma _{1}$ with a constant $\varepsilon$ $( =0.5\Omega) $, a gap is fully opened. 
Here $\lambda =0$ for both (a) and (b). 
}
\label{PT-stability}
\end{figure}

Remarkably, the present 
fully tunable experimental setup can also be exploited to examine the $PT$%
-protected topological stability of the $\mathbb{Z}_{2}$ nontrivial band
crossing points from the following aspects. First we check the topological
stability of these band crossing points of nontrivial $\mathbb{Z}_{2}$
charges. From the topological band theory, each of them should be stably
present under whatever perturbations that preserve the joint $PT$ symmetry
and do not mix one point with another, while the individual $P$ and $T$
symmetries may be violated at the same time~\cite{Zhao-Schnyder-Wang-PRL}. In this experiment, by
introducing the perturbation $H_{1}^{\prime }=\eta\sigma _{2}$ (with $%
\eta =1/2\ \Omega $ being a constant) to the system. Now the parameter of $%
\sigma _{2}$ reads $\Omega _{2}(k_{x})=\Omega (\sin k_{x}+1/2),$ which
breaks both $P$ and $T$ but preserves $PT$, it is observed that although the
band structure is distorted dramatically, and the positions as well as
neighborhood geometries of band-crossing points are changed significantly,
these band-crossing points are persistently present in the first BZ without
opening any gap, as seen clearly from Fig.\ref{PT-stability}a, being
perfectly consistent with the aforementioned facts of the topological band theory.
On the other hand, however, when another kind of perturbation $H_{2}^{\prime }(k)=\varepsilon
\sigma _{1}$ (e.g., a constant $\epsilon \sim 0.5\Omega $ ) is introduced to
the original Hamiltonian, 
$\hat{H}=\Omega /2\,\sigma _{1}+\Omega \sin k_{x}\sigma _{2}+(\lambda
\Omega +\Omega \cos k_{y})\sigma _{3}$, it is clear that the $PT$ symmetry is violated, since such
perturbations break $P$ but preserves $T$. Accordingly the topological protection, which requires the $PT$ symmetry, is discharged~\cite{Zhao-Schnyder-Wang-PRL}. In agreement with the theoretical
prediction, a trivial insulating gap is observed to be fully opened, as
shown in Fig.\ref{PT-stability}b~\cite{Notes-trivial-TI}. 

\begin{figure}[tbp]
\includegraphics[scale=0.28]{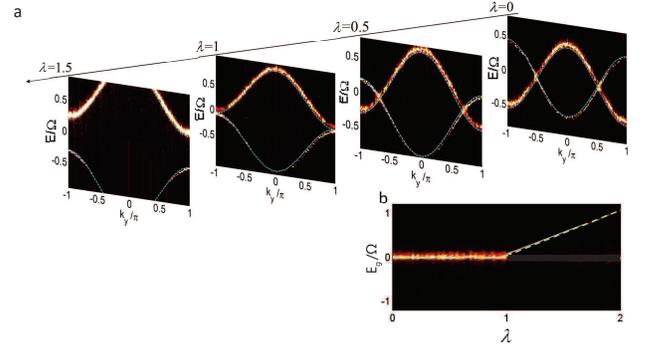} 
\caption{Quantum phase transitions from a topological gapless semimetal to a
gapped insulator as changing parameter $\protect\lambda$. a, Spectroscopy at 
$k_x\approx 0$ for various $\protect\lambda$. From right to left $\protect%
\lambda$ are $0$, $0.5$, $1$ and $1.5$, respectively. It is seen that when $%
\protect\lambda$ is increased from $0$ to $1$, then larger than $1$, the
number of Dirac-like points decreases from $4$, to $2$, then to $0$, where
the gap gradually is opened, demonstrating that a topological $PT$ invariant
semimetal phase transits to a normal insulator phase. b, Magnitude of
minimum energy gap $E_g$ in the first Brillouin zone as a function of $%
\protect\lambda$, as predicted theoretically from Eq.(\ref{Lattice-model}). }
\label{Merging-points}
\end{figure}

We now turn to examine the $\mathbb{Z}_2$ nature of the topological charge,
utilizing the fully tunable advantage of our setup. 
The spectroscopic data are shown in Fig.\ref{Merging-points}a for
representative values of $\lambda$ at each stage of the whole process of
merging and annihilation of the $\mathbb{Z}_2$ band-crossing points. 
According to general principles of topological band theory, merging two $%
\nu_{\mathbb{Z}_2}=1$ band-crossing points nucleates a band crossing point of
trivial topological charge ($\nu_{\mathbb{Z}_2}=2\equiv 0 \mod 2 $), which can be gapped
out even though the $PT$ symmetry is still preserved~\cite%
{Zhao-Schnyder-Wang-PRL}. As shown in Fig.\ref{Merging-points}b, we
continuously increase the parameter $\lambda$  
from $0$ to $2$. Starting from $\lambda=0$, where two band-crossing points
are well separated at $k_y=\pi/2$ and $-\pi/2$, respectively, in the
one-dimensional subsystem with $k_x=0$, the two band-crossing points are
gradually moving closer and closer to each other (with regard to their
distances to the BZ boundaries) when $\lambda$ is increased smoothly, then
they are merged to be a new band-crossing point at the edge of the first $BZ$
for $\lambda=1$, which should be a topologically trivial point according to
the topological band theory as mentioned above. Indeed, when $\lambda$ is
further increased to be bigger than $1$, it is observed that the band
crossing point of a trivial topological charge is gapped out, leading to a
topologically trivial insulator that has even the $PT$ symmetry~\cite{Notes-trivial-TI-2},
which verifies the aforementioned theoretical prediction.

To summarize, we have reported the first experimental realization and
manipulation of fundamental space-time inversion invariant topological
semimetal bands possessing neither $T$ nor $P$ symmetry. The non-trivial
bulk topological band structures of $PT$ symmetry have directly been imaged
with superconducting quantum circuits. Moreover, two exotic \textit{%
topological quantum phase transitions} have been observed for the first
time. The present work is expected to  stimulate a huge experimental and
theoretical interest on various $PT$ symmetric topological
metals/semimetals, paving the way for quantum-simulating novel topological
quantum materials.

\begin{acknowledgments}
\textit{Acknowledgments} This work was partly supported by the the NKRDP of
China (Grant No. 2016YFA0301802), NSFC (Grant No. 91321310, No. 11274156,
No. 11504165, No. 11474152, No. 61521001), the GRF of Hong Kong (Grants No.
HKU173055/15P and No. HKU173309/16P).
\end{acknowledgments}


\begin{thebibliography}{41}
\bibitem[{Not({\natexlab{a}})}]{Notes-PT-locality}
	\bibinfo{note}{Note that $PT$ acts locally on momentum coordinates in contrast
		to that $P$ or $T$ inverses the momentum space, which justifies the
		application of the $KO$ theory.}
\bibitem[{Not({\natexlab{d}})}]{Notes-trivial-TI}
	\bibinfo{note}{To see why this insulating phase is trivial, one may take a
		limit of the term $\Omega/2\sigma_1$ to infinity, under which the system
		stays in the same phase since there is no gap closing occurred. But the
		infinite-large limit proportional to $\sigma_1$ obviously corresponds to a
		trivial insulator.}
\bibitem[{Not({\natexlab{e}})}]{Notes-trivial-TI-2}
	\bibinfo{note}{The triviality follows from the same argument as that in
		Ref.\cite{Notes-trivial-TI}, but for the infinite limit of
		$\lambda\sigma_3$}.
\end{thebibliography}
\end{document}